\documentclass[12pt]{article}

\newcommand{\be}{\begin{equation}}
\newcommand{\ee}{\end{equation}}
\newcommand{\bea}{\begin{eqnarray}}
\newcommand{\eea}{\end{eqnarray}}

\newcommand{\bit}{\begin{itemize}}
\newcommand{\eit}{\end{itemize}}
\newcommand{\no}{\noindent}

\begin{document}

{\sf 

\title{Calibrations, Torsion Classes and Wrapped M-Branes}
\author{Ansar Fayyazuddin\footnote{email: Ansar\_ Fayyazuddin@baruch.cuny.edu} $^1$ 
and Tasneem Zehra Husain\footnote{email: tasneem@physics.harvard.edu} $^2$}

\maketitle

\begin{center}

{\it 
$^1$ Department of Natural Sciences, Baruch College, City University of New York, NY\\ 

$^2$ Jefferson Physical Laboratory, Harvard University, Cambridge, MA 02138
}

\end{center}

\vspace{1cm}

\begin{abstract}
The present work has two goals.  The first is to complete the classification of geometries in terms of torsion classes of M-branes wrapping cycles of a Calabi-Yau.  The second goal is to give insight into the physical meaning of the torsion class constraints.  We accomplish both tasks by defining new energy minimizing calibrations in M-brane backgrounds. When fluxes are turned on, it is these calibrations that are relevant, rather than those which had previously been defined in the context of purely geometric backgrounds. 
\end{abstract}

\vspace{-17cm}
\begin{flushright}
HUTP-05/A0045 \\
BCCUNY-HEP /05-05 \\
hep-th/0512030
\end{flushright}

\thispagestyle{empty}

\newpage

\tableofcontents

\section{Introduction}
In this paper we give a complete characterization of the geometry of supersymmetric backgrounds produced by wrapped branes in low-energy M-theory, a.k.a. 11-dimensional supergravity.  By "characterizing the geometry" we will mean specifying the torsion classes of these geometries as defined in \cite{CS} and further explored in the string theory context in \cite{torsionclasses}.  In the following we will study branes wrapping cycles in Calabi-Yau 3-folds.

We have two distinct goals.  On the one hand we would like to rectify the prevailing incomplete knowledge of wrapped brane geometries, focusing particularly on specifying the torsion classes of such geometries completely, where that knowledge is lacking. The second goal is to gain a real understanding for the physical meaning of the torsion class constraints. 

Our approach is based on calibrations.  Calibrations in purely geometric backgrounds give a lower bound on the volume of certain types of cycles. Branes wrapped on these minimal volume cycles also turn out to have minimal energy (mass). This latter observation, first used in \cite{spells2} helps us to extend the concept of calibrations to more general supergravity backgrounds, where fields other than the metric are turned on; we thus define calibrations in such a way that they give us lower bounds on mass. In keeping with this logic, we will compute the mass of supersymmetric configurations in the supergravity background of wrapped branes. Since we find that the mass is given by the integral of certain forms, we will then require that these be calibrating forms and exploit that property that to arrive at constraints on the underlying supergravity background.  We explain our methods in more detail below.      

\section{Calibrations and Torsion Classes}
In this section we will explore the geometry produced by wrapped M-branes in Calabi-Yau 3-folds.  We will study this geometry at three different levels.  At the first level we consider M-theory compactified on a Calabi-Yau 3-fold.  Space-time is thus given by the product of 5 dimensional Minkowski space and the Calabi-Yau manifold:
\be
ds^2 = \eta_{\mu\nu}dx^{\mu}dx^\nu + 2g_{M\bar{N}}dz^Mdz^{\bar N}
\ee
The Calabi-Yau manifold has non-trivial 2-cycles, 4-cycles, and 3-cycles in addition to a 0 and 6-cycle.  The 2-, 4- and 6-cycles are holomorphic cycles calibrated by $J$, $J \wedge J$ and $J \wedge J \wedge J$ respectively, where $J$ is the K{\" a}hler form on the CY3, and the 3-cycles are Special Lagrangian cycles calibrated by the unique (3,0) form.  

At the second level we study the geometry produced by wrapped branes.  The constraints on supersymmetric solutions of supergravity describing M2 and M5-branes wrapping different cycles of Calabi-Yau 3-folds have been studied \cite{bfhs, Tasneem, m&s, kastor, amherst}. There are some universal properties shared by all such solutions, and we now proceed to collect these.  Let us suppose that we have an Mp-brane wrapping a d-cycle with $d\leq p$ of the Calabi-Yau.  Then the geometry is such that:
\be
ds^2 = H_1^2\eta_{\mu\nu}dx^\mu dx^\nu + g_{IJ}dy^Idy^J + H^2_2dx_\perp^2 \label{maing}
\ee
where the $y$ coordinates are on a manifold we will call $M$. Even though this is the part of the space that was originally a Calabi-Yau, it has now been modified by of the presence of the wrapped Mp-brane.  The $x$ coordinates originally spanned the 5-dimensional Minkowski space transverse to the Calabi-Yau. They are further divided into a warped Minkowski part of dimension $p-d+1$ labeled by $x^\mu$, and another flat Euclidean part transverse to the brane $x_\perp$ of dimension $4-p+d$. In what follows, we will take all branes to be located at $x_\perp =0$.  

There are some important general facts about the manifold $M$ on which the $y^I$ are coordinates.  The manifold has an almost complex structure - i.e. the cotangent space is a direct sum of globally defined (1,0) and (0,1) forms which are related to each other by complex conjugation.  We can take a convenient basis of (1,0) forms to be $e^m, m=u,v,w$ which provide a local frame satisfying $g_{IJ} =\eta_{m\bar{n}}(e^m_Ie^{\bar n}_J + e^m_Je^{\bar n}_I)$.  The existence of an almost complex structure implies that a $U(3)$ structure exists on the manifold.  One consequence of this fact is that the (3,0) form $\Omega = e^u\wedge e^v\wedge e^w$ is {\it not} a priori globally well defined because it could, in principle, pick up a phase under the U(1) rotations which are now possible.  We will see, however, that it {\it does} turn out to be well defined and therefore we are justified in saying that what in fact exists is a reduced SU(3) structure.  

The geometry of wrapped M-branes can be classified through torsion classes \cite{CS, torsionclasses}.  Briefly the torsion classes are specified through a (1,1) form $J \equiv \frac{1}{2}(e^u\wedge e^{\bar u} + e^v\wedge e^{\bar v} + e^w\wedge e^{\bar w})$ and a (3,0) form $\Omega$.  These forms satisfy the identities:
\bea
\Omega\wedge J &=&0 \nonumber \\
\Omega\wedge \bar{\Omega} &=& \frac{4}{3}J\wedge J\wedge J
\eea
The torsion classes of a particular manifold are specified by using the almost complex structure to classify the derivatives of the corresponding $J$ and $\Omega$ as $(p,q)$ forms.

The purpose of this paper is to investigate the geometry (\ref{maing}) for different M-brane configurations.  We will work at a new {\it third} level to investigate (\ref{maing}).  At this level, we introduce new supersymmetric wrapped branes in the {\it background} (\ref{maing}) and ask what their mass is.  For these branes we do not need to find the "backreaction" on the geometry; they are introduced only as probes whose energy can be computed.  We will always require that the probe Mq-branes have worldvolume such that they wrap a $d'$ cycle in the Calabi-Yau with the remaining directions of the Mq-brane entirely contained inside of the warped Minkowski space.  In other words it will be important to us that $q-d' < p-d$.  With this requirement the probe wrapped brane is a (q-d')-brane in the (p-d+1)-dimensional worldvolume theory of the background Mp-brane.  The mass of the (q-d')-brane is then an object whose mass is a meaningful quantity in the worldvolume theory which has $x^0$ as its time coordinate.

More concretely, consider an Mq-brane wrapping a cycle $\Sigma_d$ in the Calabi-Yau.  The action of the Mq-brane in the background geometry (\ref{maing}) is:
\be
S = \tau_q\int H_1^{q-d'+1}d\mbox{vol}(\Sigma_d')\wedge dx^0\wedge ...\wedge dx^{(q-d')}
\ee
where $\tau_q$ is the (constant) 11-d tension of the Mq-brane and $dvol(\Sigma_d')$ is the volume form on $\Sigma_d$ induced by the metric (\ref{maing}).  The Mq-brane appears as a $(q-d')$-brane in the warped Minkowski part of the worldvolume of the background Mp-brane.  The tension of this $q-d'$ brane is:
\be
T_{(q-d')} = \int H_1^{q-d'+1}d\mbox{vol}(\Sigma_d'). \label{tension}
\ee
In general such a brane configuration is un-stable and will decay to a lower energy configuration.  If we require the probe Mq-brane to be supersymmetric, however, stability is guaranteed.  It is easy to show that for supersymmetric probe Mq-brane configurations
\be
d\mbox{vol}(\Sigma_d') = \phi_{d'} |_{\Sigma_{d'}}.\label{minvolume}
\ee
where $\phi_{d'}$ is an appropriate $d'$-form defined on $M$ - the 6-d part of the space-time which was once the Calabi-Yau, before the effects of the background Mp-brane were taken into account.  This suggests that there is a calibration associated with the tension of the $(q-d')$-dimensional object in the $(p-d+1)$-dimensional theory on the part of the background Mp-brane which is transverse to the Calabi-Yau.  This calibration can be read off from (\ref{tension}) taking into account (\ref{minvolume}):
\be
\psi_{d'} = H_1^{q-d'+1}\phi_{d'}.
\ee
So, $\psi_{d'}$ is a calibrating $d'$-form.  Since calibrations give us a lower bound on the mass, they must only care about the topological class of what they are integrated over and are necessarily closed.  Thus supersymmetry requires that $\psi_{d'}$ be closed:
\be
d_6\psi_{d'} =0 \label{mastereq}.
\ee

It is straightforward to find all possible $\phi_{d'}$.  Since we are dealing here with Calabi-Yau 3-folds, the only non-trivial values $d'$ can take are $d'= 2,3,4$ corresponding to the three different non-trivial Hodge numbers of Calabi-Yau 3-folds.  In the background geometry (\ref{maing}) the supersymmetry preserved by a probe Mq-brane is given by the projection:
\be
\partial_{\alpha_1}Y^{I_1}...\partial_{\alpha_{d'}}Y^{I_{d'}}\Gamma_{0...(q-d')}\Gamma_{I_1... I_{d'}}\eta = H_1^{(q-d'+1)/2}dvol(\Sigma_{d'})\eta.
\ee
It is easy to see that to preserve supersymmetry M5-branes can intersect other M5-branes along either $2$ or $4$ space-time directions, M5-branes can also intersect M2-branes along $2$ space-time dimensions, while, in addition to the intersections already encountered, M2-branes can intersect other M2-branes along $1$ space-time directions.  In all of these cases one can easily establish that:
\bea
\phi_2 &=& J \nonumber \\
\phi_3 &=& \Omega \\
\phi_4 &=& J\wedge J \nonumber 
\eea
 
Our strategy will be to take different background Mp-branes and introduce all possible probe Mq-branes.  Using the above considerations we derive equations (\ref{mastereq}) to find constraints on the background geometry of the Mp-brane.  The ideas outlined here were first used in \cite{spells2} to interpret  a new condition on a (2,0) form which arose as an extension of the results of \cite{danda}.  
    
\section{Torsion classes from calibrations}

In this section we apply the formalism developed in the previous section.  We consider background geometries of M5- and M2-branes wrapping different cycles of Calabi-Yau manifolds.  We will introduce all possible supersymmetric probes and derive constraints on the background geometry from requiring that the mass calibrating form (\ref{mastereq}) $\psi_{d'}$ for the probe is closed.  In what follows we use the convention that for wrapped M5 backgrounds $H_1^2 = H^{-1/3}$ and for wrapped M2-branes $H_1^2 = H^{-2/3}$.

\subsection{M5-brane on a holomorphic 2-cycle}

Consider an M5-brane wrapped on a holomorphic 2-cycle inside a Calabi-Yau 3-fold.  According to the supersymmetric intersection rules outlined in the previous section, we can introduce an M5-brane wrapping a 4-cycle or an M5-brane wrapping a Special Lagrangian 3-cycle, while still preserving supersymmetry since all these probe branes satisfy the rules outlined in the previous section.

From here onwards, we will lay out the brane configurations in tabular form to make the set-up more transparent.  The brane which gives rise to the background will have world-volume directions denoted by $\otimes$, and the probe branes we introduce will have world-volumes in the $\times$ directions.  

If the CY 3-fold spans directions $456789$, then an M5-brane wrapping a holomorphic 2-cycle 
can schematially be represented as follows:

\be
\begin{array}[h]{|ccccc|cccccc|c|}
  \hline
   \; & 0 & 1 & 2 & 3 & 
              4 & 5 & 6 & 7 & 8 & 9 & 10\\
  \hline
  {\bf M5} & \otimes & \otimes & \otimes & \otimes & \otimes & \otimes  
               &  &  &  &  & \\
  \hline
\end{array}
\ee

\no
We first introduce an M5-brane probe which wraps a SpelL 3-cycle in what was once the Calabi-Yau. This brane will be BPS only if it shares 3-spatial directions with the previous 5-brane, i.e, it should correspond to something like this:

\be
\begin{array}[h]{|ccccc|cccccc|c|}
  \hline
   \; & 0 & 1 & 2 & 3 & 
              4 & 5 & 6 & 7 & 8 & 9 & 10\\
  \hline
  {\bf M5} & \otimes & \otimes & \otimes & \otimes & \otimes & \otimes  
               &  &  &  &  & \\
{\bf M5} & \times & \times & \times & & \times & &   
             \times  &  & \times &  &  \\
  \hline
\end{array}
\ee

\no
Since we have ensured that this probe is supersymmetric, we know that its mass must 
correspond to an integral of a calibrating form in the geometry 
we are probing. In other words, the mass density must be a closed form. Hence, we find that 
\be
\psi_3 = H^{-1/2} \Omega
\ee
is closed with respect to $d_6$ the exterior derivative on $M$:
\be
d_6(H^{-1/2} \Omega) = 0
\ee
As far as we are aware this equation has not appeared previously in the literature and is therefore a new result.
\no

The same background can also be probed by an M5-brane wrapping a holomorphic 4-cycle. This would correspond to the following type of set-up. 
\be
\begin{array}[h]{|ccccc|cccccc|c|}
  \hline
   \; & 0 & 1 & 2 & 3 & 
              4 & 5 & 6 & 7 & 8 & 9 & 10\\
  \hline
  {\bf M5} & \otimes & \otimes & \otimes & \otimes & \otimes & \otimes  
               &  &  &  &  & \\
{\bf M5} & \times & \times & & & \times & \times & \times & 
             \times  &  & &   \\
  \hline
\end{array}
\ee

\no
Applying our formalism from the previous section, we find that $\psi_4 = H^{-1/3}J\wedge J$ and therefore,
\be
d_6(H^{-1/3}J\wedge J) = 0
\ee
This equation was first derived in \cite{bfhs} by solving the Killing spinor equation for the background.  Here we derive it in a very simple way and give the equation a physical meaning - as arising from calibrating the mass of a probe M5-brane wrapping a 4-cycle.  

\subsection{M5-brane on a SpelL 3-cycle}

We now move on to an M5-brane wrapping a Special Lagrangian 3-cycle in a 
Calabi-Yau 3-fold, spanning directions $345678$.  This problem was studied in \cite{m&s, spells}

\be
\begin{array}[h]{|cccc|cccccc|cc|}
  \hline
   \; & 0 & 1 & 2 & 3 & 
              4 & 5 & 6 & 7 & 8 & 9 & 10\\
  \hline
  {\bf M5} & \otimes & \otimes & \otimes & \otimes & & \otimes & & \otimes  
               &  &  & \\
  \hline
\end{array}
\ee

\no
An M5-brane probe wrapping a holomorphic 4-cycle can be introduced into this background, 
such that the resulting configuration is still BPS. 

\be
\begin{array}[h]{|cccc|cccccc|cc|}
  \hline
   \; & 0 & 1 & 2 & 3 & 
              4 & 5 & 6 & 7 & 8 & 9 & 10\\
  \hline
  {\bf M5} & \otimes & \otimes & \otimes & \otimes && \otimes && \otimes  
               &  &  &  \\
{\bf M5} & \times & \times & & \times & \times &  \times & 
             \times  &  & & &   \\
  \hline
\end{array}
\ee

\no
The volume along of the 4-cycle is given by $J\wedge J$ and by the considerations outlined in the previous section the mass density is given by:
\be
\psi_4 = H^{-1/3} J \wedge J
\ee
and so we find that 
\be
0 = d_6 \psi_4 = d_6(H^{-1/3} J \wedge J)  \label{spell1}
\ee

In addition, we can introduce a probe instanton - a membrane that wraps a SpelL 3-cycle; since this probe is a Euclidean brane, it must now share two worldvolume directions with the M5-brane.
If we stick with the convention that the M5-brane is wrapping the cycle calibrated by Re $\Omega$, then the volume of the Special Lagrangian cycle wrapped by the M2-brane is given by the pullback of $Im\Omega$. 
\be
\begin{array}[h]{|cccc|cccccc|cc|}
  \hline
   \; & 0 & 1 & 2 & 3 & 
              4 & 5 & 6 & 7 & 8 & 9 & 10\\
  \hline
  {\bf M5} & \otimes & \otimes & \otimes & \otimes & &\otimes & &\otimes  
               &  &  &  \\
{\bf M2} & & & & \times & & \times & & &\times 
             & &   \\
  \hline
\end{array}
\ee
 Since the M2-brane has no space-time directions along the $0,1,2$ directions the form
\be
\psi_3 = Im\Omega
\ee
calibrates the action of the M2-brane instanton, hence
\be
d_6 Im\Omega = 0 \label{spell2}
\ee
Finally, we can introduce an M2-brane wrapping a holomorphic 2-cycle in the Calabi-Yau:
\be
\begin{array}[h]{|cccc|cccccc|cc|}
  \hline
   \; & 0 & 1 & 2 & 3 & 
              4 & 5 & 6 & 7 & 8 & 9 & 10\\
  \hline
  {\bf M5} & \otimes & \otimes & \otimes & \otimes & &\otimes & &\otimes  
               &  &  &  \\
{\bf M2} &\times & & & \times &\times &  & & & 
             & &   \\
  \hline
\end{array}
\ee
In this case the volume of the M2-brane inside the Calabi-Yau is given by $J$ so that 
 \be
 \psi_2 = H^{-1/6}J
\ee
which is again closed:
 \be
 d_6\psi_2 = d_6(H^{-1/6}J) =0 \label{spell3}
\ee
All of these equations, i.e. (\ref{spell1}), (\ref{spell2}), (\ref{spell3}), are known from \cite{m&s, spells}.  We have provided a physical way of understanding why these otherwise mysterious conditions arise.

\subsection{M5-brane on a holomorphic 4-cycle}

Finally, we turn to the background generated by an M5-brane on a holomorphic 4-cycle inside a Calabi-Yau 3-fold spanning
directions $234567$

\be
\begin{array}[h]{|ccc|cccccc|ccc|}
  \hline
   \; & 0 & 1 & 2 & 3 & 
              4 & 5 & 6 & 7 & 8 & 9 & 10\\
  \hline
  {\bf M5} & \otimes & \otimes & \otimes & \otimes & \otimes & \otimes  
               &  &  &  &  & \\
  \hline
\end{array}
\ee

\no
In this case we come up against our constraint that the part of the probe brane transverse to the Calabi-Yau lie entirely inside of the background brane (our constraint that $p-d>q-d'$).  This constraint prevents us from wrapping an M-brane on a holomorphic cycle while preserving supersymmetry.   Thus we cannot use our methods to find constraints on $J$ or $J\wedge J$.  Fortunately we already know from previous work \cite{amherst, Tasneem} that the constraint satisfied by $J$ is $d_6(H^{1/3}J)=0$.  We don't, however, know anything about $\Omega$.  Fortunately we can deduce its properties using our methods.  We introduce for this purpose an M2-brane instanton wrapping a Special Lagrangian 3-cycle in the Calabi-Yau.    

\be
\begin{array}[h]{|ccc|cccccc|ccc|}
  \hline
   \; & 0 & 1 & 2 & 3 & 
              4 & 5 & 6 & 7 & 8 & 9 & 10\\
  \hline
  {\bf M5} & \otimes & \otimes & \otimes & \otimes & \otimes & \otimes  
               &  &  &  &  & \\
{\bf M2} &  &  & & \times &  & \times &  & \times  
               &  &  &   \\
  \hline
\end{array}
\ee

Once again, since the M2-brane is entirely inside of the Calabi-Yau there are no factors of $H_1$ in our definition of $\psi_3 = \Omega$ and we have the constraint:
\be
d_6\Omega = 0.
\ee 
As far as we know this is also a new result.

\subsection{M2-brane on a holomorphic 2-cycle}

We turn now to the M2-brane as it wraps a holomorphic 2-cycle,

\be
\begin{array}[h]{|cc|cccccc|cccc|}
  \hline
   \; & 0 & 1 & 2 & 3 & 
              4 & 5 & 6 & 7 & 8 & 9 & 10\\
  \hline
  {\bf M2} & \otimes & \otimes & \otimes & & &  
               &  &  &  &  & \\
  \hline
\end{array}
\ee

Once again we cannot use our methods to constrain $J$ since there are is no way to wrap M-branes on holomorphic cycles while preserving both supersymmetry and our constraint that $p-d > q-d'$ in the notation of section 2.  Again we are fortunate that the constraints are already well known and can be stated simply as \cite{Tasneem}: $d_6[H^{1/3} J \wedge J] = 0$.  

We can, however, find constraints on $\Omega$ by wrapping an M2-brane on a Special Lagrangian 3-cycle as follows, say,:
\be
\begin{array}[h]{|cc|cccccc|cccc|}
  \hline
   \; & 0 & 1 & 2 & 3 & 
              4 & 5 & 6 & 7 & 8 & 9 & 10\\
  \hline
  {\bf M2} & \otimes & \otimes & \otimes & & &  
               &  &  &  &  & \\
{\bf M2} & & \times & & \times & & \times & &  
                &  &  & \\
  \hline
\end{array}
\ee

This is a supersymmetric configuration and yields $\psi_3 = \Omega$, or the constraint 
\be 
d \Omega = 0
\ee
This is a new constraint which has not previously appeared in the literature to our knowledge.
\\

\section{Summary and new directions}
In this paper we have derived a number of {\it new} results as well as given a physical understanding of some known results.  We argued that it is sensible to define calibrations $\psi_{d}$ in wrapped M-brane backgrounds.  These calibrations measure the tension of supersymmetric objects in the worldvolume theory of the background wrapped M-brane.  We showed that many of the torsion class constraints on the geometry can be understood as the constraint that our calibrations $\psi_d$ are closed with respect to the exterior derivative on $M$.  

In our approach we do not have to satisfy the gravitino Killing spinor equation explicitly to find constraints on the metric of the wrapped brane.  We make only two assumptions: that there is an almost complex structure defined on the six dimensional manifold $M$, and that the metric takes the generic form (\ref{maing}).  With these two simple conditions we introduce all possible M-branes which intersect the background brane so that the intersection is a supersymmetric object in the background brane's worldvolume theory.  We then evaluate the action of the probe brane in the putative background of the wrapped brane and deduce facts about the metric (\ref{maing}).  We are able to deduce already known results, which gives us confidence in our methods, as well as new results.  

Our new results all involve properties of the (3,0) form $\Omega$ in backgrounds  of M-branes wrapping {\it holomorphic} cycles.  In all of these cases we find that $d_6(H^\alpha\Omega) =0$ for some power $\alpha$.  This tells us that $d_6\Omega$ is a (3,1) form.  The fact that $d_6\Omega$ has no (2,2) component means that the almost complex structure is integrable.  This is our first important new result because it justifies {\it ex post facto} the assumptions in \cite{bfhs, amherst, Tasneem} that $M$ is a complex manifold.  Moreover, we find that for M5-branes wrapping a 2-cycle $d_6(H^{-1/2}\Omega) = 0$, while for M5-branes wrapping a 4-cycle and M2-branes wrapping 2-cycles $d_6\Omega =0$.  This new information combined with results from \cite{bfhs, amherst, Tasneem} completes their description in terms of torsion classes.  

One may wonder if there are other areas where one can apply the ideas developed here.  Clearly there is no need to have a background M-brane to apply our methods.  In fact, one can re-discover properties of Calabi-Yau manifolds if one starts out assuming that there exists an almost complex structure and introducing supersymmetric probe wrapped branes.  One finds that Calabi-Yau manifolds are complex K{\"a}hler manifolds with a holomorphic (3,0) form!  Another possibility is to apply these ideas to the geometry of branes wrapping more than one type of cycle.  For instance one could find constraints on the geometry of an M5-brane wrapping a 2-cycle and another one wrapping a 4-cycle.  These cases have not been studied before and can be treated in a very elementary way.

The basic ideas developed here can be generalized to other manifolds and branes as well.  
\\
\\
 
\no
{\Large \bf Acknowledgements}\\

\no

AF would like to thank VR for funding.  TZH is grateful to Sergei Gukov for the fascinating discussion that was the seed of this project and would like to acknowledge funding from VR. We are grateful to the NYC West Village cafe {\it Grounded}, for the inspiring setting and their patience with our extended sojourn!

}
\end{document}